\documentclass[prl,twocolumn,showpacs,preprintnumbers,amsmath,amssymb]{revtex4}

\usepackage{graphicx}
\usepackage{dcolumn}
\usepackage{bm}

\begin{document}

\preprint{}

\title{Macroscopic quantum tunneling in a $d$-wave high-$T_{C}$ Bi$_{{\rm 2}}$Sr$_{{\rm 2}}$CaCu$_{{\rm 2}}$O$_{{\rm 8}{\rm +} {\rm \delta} }$ superconductor}

\author{K. Inomata$^{1, 2}$}
\author{S. Sato$^{1}$}
\author{Koji Nakajima$^{1}$}
\author{A. Tanaka$^{2}$}
\author{Y. Takano$^{2}$}
\author{H. B. Wang$^{2}$}
\author{M. Nagao$^{2}$}
\author{H. Hatano$^{2}$}
\author{S. Kawabata$^{3}$}
\affiliation{$^{1}$Research Institute of Electrical Communication
(RIEC), Tohoku University, Sendai
980-8577, Japan}

\affiliation{$^{2}$National Institute for Materials
Science (NIMS), Tsukuba 305-0047, Japan}

\affiliation{$^{3}$National Institute of Advanced Industrial Science and
Technology (AIST), Tsukuba 305-8586, Japan }

\begin{abstract}
While Josephson-junction-like
structures intrinsic to the layered cuprate high temperature
superconductors offer an attractive stage for exploiting possible
applications to new quantum technologies, the
low energy quasiparticle excitations characteristically present in
these $d$-wave superconductors may easily destruct the coherence
required. Here we demonstrate for the first time the feasibility
of macroscopic quantum tunneling in the intrinsic Josephson
junctions of a high temperature superconductor Bi$_{{\rm 2}}$Sr$_{{\rm
2}}$CaCu$_{{\rm 2}}$O$_{{\rm 8}{\rm +} {\rm \delta} }$, and find
it to be characterized by a high classic-to-quantum crossover
temperature and a relatively weak quasiparticle dissipation.
\end{abstract}

\maketitle

A marked feature characterizing cuprate high temperature
superconductors (HTSC) is its strong two dimensionality. In
particular the bismuth-based HTSCs, for which this anisotropy is
prominent, are best viewed as stacks of superconducting CuO$_{{\rm 2}}$
planes weakly linked through intrinsic Josephson junction
(IJJ) type couplings \cite{Kleiner:1992}. These built-in atomic
scale links are taylor-made for technical applications difficult
to achieve with artificial Josephson junctions (JJ), and its study
has now developed into an active interdisciplinary field. We
report below what is to our knowledge the first successful
observation of the macroscopic quantum tunnelling (MQT)
\cite{Caldeira:1981,Voss:1981,Clarke:1988,Devoret:1985} of the
phase variable of the superconducting order parameter through the
potential barrier of an IJJ, opening up an entirely new direction
for HTSC applications. While the corresponding phenomena had been
observed at around 300 mK in conventional JJs
\cite{Voss:1981,Devoret:1985,Wallraff:2003}, we have confirmed MQT
behavior at approximately 1 K apparently reflecting the
characteristically high plasma frequency of IJJs. Our results are
highly nontrivial in that they also demonstrate the feasibility of
MQT in spite of the presence of dissipative low energy
quasiparticles
\cite{Fominov:2003,Amin:2004,Joglekar:2004,Kawabata:2004}, which
is the other hallmark of HTSCs.

Current-biased JJs offer an ideal stage for realizing a variety of
macroscopic quantum phenomena, e.g. energy level quantization
within the potential well
\cite{Clarke:1988,Martinis:1985,Silvestrini:1997} and the
associated MQT and macroscopic quantum coherence
\cite{Mooij:1999,Caspar:2000}, all of which have come to be
recognized as having immediate implications for qubit applications
\cite{Mooij:1999,Martinis:2002,Han:2002,Nakamura:1999}. In
particular, a phase qubit utilizing MQT has been reported by
Martinis and co-workers \cite{Martinis:2002}.

Aside from external noises and disorder, a primary source which
stands as an obstacle towards observation of MQT is the influence
of non-superconducting quasiparticle excitations
\cite{Caldeira:1981}. In conventional $s$-wave superconductors,
all quasi-particle states are separated from the superconducting
ground state by a finite energy gap and thus become essentially
inaccessible upon going to sufficiently low temperatures; hence
the observability of MQT. The situation is drastically altered
when we turn to HTSCs, which are $d$-wave superconductors. The
latter are characterized by four nodes in the order parameter at
which the energy gap vanishes \cite{Harlingen:1995,Tsuei:2000}.
HTSCs therefore necessarily sustain a finite population of
dissipative quasiparticles down to the lowest temperatures, which
would lead one to render MQT unattainable. While this accounts for
the absence of previous phase-MQT reports in HTSCs, several
theoretical works have recently put this naive view into question
by inferring that dissipation due to nodal quasiparticles is not
strong enough to totally destruct the coherence required for MQT
\cite{Fominov:2003,Amin:2004,Joglekar:2004,Kawabata:2004}.
In addition to the moderately weak influence
of quasiparticles, it is also crucial from the experimental point
of view to have (\ref{eq1}) JJs bearing a high degree of
underdamping and hysterics, and (\ref{eq2}) a measurement system
sufficiently detached from the environment. Devising samples and
apparatus fulfilling this prerequisite constituted an integral
portion of our project of observing MQT in the IJJs of Bi$_{{\rm
2}}$Sr$_{{\rm 2}}$CaCu$_{{\rm 2}}$O$_{{\rm 8}{\rm +} {\rm \delta}
}$ (Bi-2212).

We now turn to the description of our samples and measurement
set-up. In Bi-2212, it is well known that the stacking crystal
structure along the $c$ axis consists of an alternating array of
the superconducting (S) CuO$_{{\rm 2}}$ layers and the
insulating (I) BiO and SrO layers. The excellent JJ properties
exhibited by IJJs in Bi-2212 is by now well established; being
ideal atomic sized junctions \cite{Kleiner:1992,Kleiner:1994},
they are in particular free of defects and surface roughness of
the insulating oxide layers which degrade the quality of
artificially made JJs, such as in Al and Nb. To probe the
characteristics of the built-in IJJs, it is necessary to construct
a pass for the interlayer current flow as illustrated in Fig. 1(a)
employing the combination of the double side etching and 3-D
focused ion-beam etching techniques \cite{Wang:2001,Kim:1999} .
The junction thus fabricated will generally consist of several
layers of IJJs coupled inductively and via charging effects --the
number of which is technically difficult to control owing to its
minute (Angstrom) scale. Therefore, the IJJs form a 1D-array of
S-I-S type JJ. Close to zero bias, the behavior is nevertheless
well captured in terms of a single junction model
\cite{Machida:1999}, as evidenced from our experiments detailed
below. The IJJs were mounted on the copper-mixing chamber of a
$^{{\rm 3}}$He/$^{{\rm 4}}$He Oxford dilution refrigerator with an
excellent thermal contact and cooled down from 4.2 K and 40 mK.
All cable connections at room temperature were guided through high
impedance cupronickel semi-rigid coaxial cables, which work as
a low-pass filter above 1 GHz, to avoid propagating thermal
flows and external noises into the sample. Moreover, by using
\textit{LC} low-pass filters along these cables we achieved a
strong attenuation of the external noises above 15 MHz. The
dilution refrigerator along with the entire analog-measurement
system was placed in a shielding room.

\begin{figure}
\includegraphics[width=8.2cm]{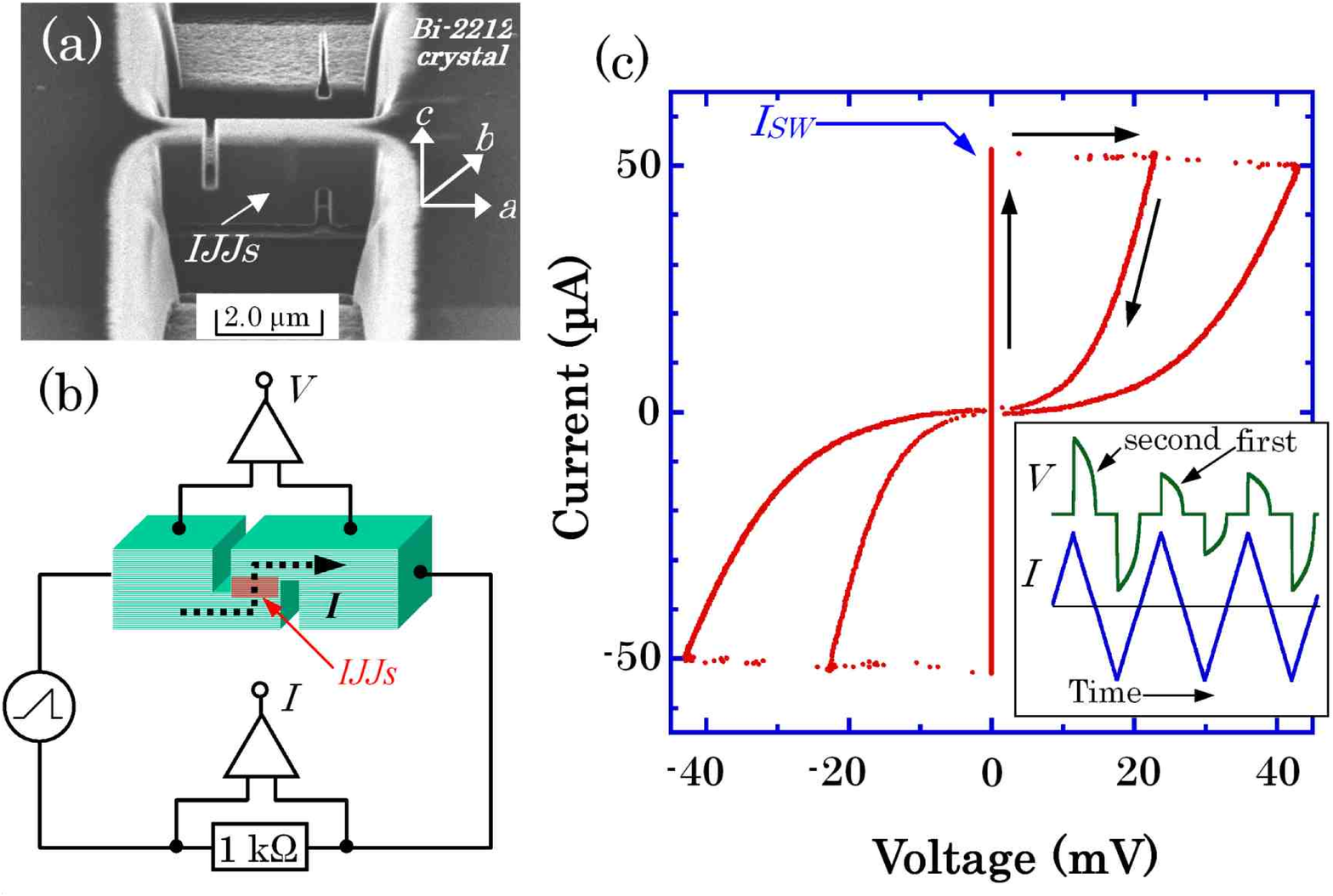}
\caption{\label{fig.1} (a) Scanning ion-beam micrograph of the
IJJs. The critical temperature of the sample is around 90 K. The
number of elemental junctions in the IJJs is less than 20 and the 
junction dimension in the $ab$ plane is 1.76 $\times $ 0.86 $\mu$m$^{{\rm 2}}$.
(b) Schematic of the bias lead configuration. Electrical properties
of IJJs were measured with the four probes method. (c) $I$-$V$ characteristic
of the IJJs measured at 4.2 K. The $I$-$V$ curves were measured by ramping
a bias current up and down repeatedly. A total of two resistive branches
(first and second branches) are exhibited in this figure. The inset shows
$V$ or $I$ vs. time.}
\end{figure}

The dynamics of a Josephson junction
\cite{Barone:1982,Tinkham:1996} is conveniently described as the
motion of a fictitious particle in a washboard potential $U(\phi
)=-U_{0}(x\phi +\cos\phi )$, where the phase difference $\phi $ is
to be viewed as the particle coordinate. Here $U_{0}=\hbar
I_{C}/2e$ is the Josephson coupling energy and $x=I/I_{C}$ ($I$:
bias current, $I_{C}$: critical current). The particle at the
bottom of the potential well oscillates at the plasma frequency
defined as $\omega _{p}=\omega _{p0}(1-x^{{\rm 2}})^{{\rm 1}{\rm
/} {\rm 4}}$, where $\omega _{p0}=(2eI_{C}/\hbar C)^{{\rm 1}{\rm
/} {\rm 2}}$ is the zero-bias plasma frequency and $C$ is the
junction capacitance. In the presence of a bias-current, $U(\phi
)$ consists of a periodic potential structure separated by the
energy barrier $E(x)=U_{0}\{-\pi x+2[x\sin^{{\rm -} {\rm
1}}x+(1-x^{{\rm 2}})]^{{\rm 1}{\rm /} {\rm 2}}\}$. Thermal
fluctuations at sufficiently high temperatures assists a
thermally-activated escape of the particle across the barrier
whose rate is given by \cite{Fulton:1974}
\begin{equation}
\label{eq1} \tau _{TA} ^{ - 1} = {\frac{{\omega _{p}} }{{2\pi}
}}\exp \left( { - {\frac{{E(x)}}{{k_{B} T}}}} \right){\rm ,}
\end{equation}
where $k_{B}$ is the Boltzmann constant and $T$ is the
temperature. As $T$ is lowered, thermal activation is suppressed
exponentially and the escape process eventually gets dominated by
MQT events. This crossover takes place at the temperature
$T^{*}\sim \hbar \omega _{p}/2\pi k_{B}$
\cite{Silvestrini:1997,Grabert:1984}, and the MQT rate is
expressed as \cite{Caldeira:1981}
\begin{equation}
\label{eq2} \tau _{MQT} ^{ - 1} = 12\omega _{p} \left(
{{\frac{{3E(x)}}{{2\pi \hbar \omega _{p}} }}} \right)^{1 / 2}\exp
\left( { - {\frac{{36E(x)}}{{5\hbar \omega _{p}} }}} \right){\rm
.}
\end{equation}

The current-voltage ($I$-$V$) characteristic of the IJJs measured
by biasing along the $c$ axis (see Fig. 1(b)) at 4.2 K is
displayed in Fig. 1(c). The typical $I$-$V$ characteristic of IJJs
exhibiting a multi-branch structure was observed. The total number of
branches corresponds to the number of elemental junctions in the IJJs
as one can confirm by applying a larger bias. When biased, a
switching to a finite voltage state takes place at a random value
of the switching current $I_{SW}$ and the $I$-$V$ curve shows a
hysteresis. The zero-bias Josephson current is slightly larger
than the first branch of IJJs, which indicates that the noise
level of our system is quite low. In addition, the McCumber
parameter $\beta _{C}$=$2eI_{C}CR_{q}^{{\rm 2}}/\hbar$ and the
damping parameter $Q^{{\rm -} {\rm 1}}$=$(\omega _{p0}CR)^
{{\rm -} {\rm 1}}$ were $\sim
10^{{\rm 5}}$ and $\sim 10^{{\rm -} {\rm 2}}$, respectively.
$R$ is the effective shunt resistance, and $R_{q}$ is the quasiparticle
resistance which is estimated
from the slope of the first branch of an $I$-$V$ curve at zero bias.
This high degree of hysteresis and the underdamping are crucial for
detecting the MQT.

The physical nature of the escape process in JJs can be elucidated
by analyzing the statistics of the stochastic switching
\cite{Wallraff:2003,Silvestrini:1997,Fulton:1974} from a
zero-voltage state to a finite voltage state. To this end
measurement of $I_{SW}$ was repeated for 2000 times at each
temperature with a current resolution $\Delta I$ of 20 nA,
which we checked were sufficient to determine the switching
current distribution $P(I_{SW})$. Figure 2(a) shows the
temperature dependence of $P(I_{SW})$. The measurements were
performed for the first branch of IJJs. In this experiment a
ramp-shaped waveform was used with the constant bias speed
$dI/dt$. Histograms of $P(I_{SW})$ were calculated against the
bias current and the bin width corresponds to the current
resolution $\Delta I$ of the measurement setup. As $T$ is lowered,
$P(I_{SW})$ narrows and shifts to higher currents. The $P(I_{SW})$
measured at 4.2 K are fitted with the thermal activation theory
proposed by Kramers \cite{Kramers:1940,Garg:1995}, and the
zero-noise critical current obtained from the fitting is
$I_{C}$=48.54$\pm $0.02 $\mu $A. This value gives a practical
estimate of $I_{C}$ at 0 K. The results exhibit good agreements
with the Kramers theory at $4.2\sim1.1$ K, but deviates
substantially below 1K. Note also that below 1K, the amount of
shift of $P(I_{SW})$ along the current axis reduces considerably.
These findings strongly infer that quantum tunneling takes over
and the escape process is essentially $T$-independent in this
regime as verified below. We also expect that the good fitting of
$P(I_{SW})$ in the thermally activated regime to conventional
theories is an indicator that the weakest link in our IJJs
effectively behaves like a single Josephson junction. Figure 2(b)
shows $\tau ^{{\rm -} {\rm 1}}(I_{SW})$ calculated from the
experimental data in Fig. 2(a)
\cite{Silvestrini:1997,Fulton:1974}. As $T$ is lowered, the slope
of $ln$ $\tau ^{{\rm -} {\rm 1}}(I_{SW})$ increases and $\tau
^{{\rm -} {\rm 1}}(I_{SW})$ shifts to higher $I_{SW}$ similar to
the behavior of $P(I_{SW})$. One observes from a comparison of the
two figures that below 1 K, where the amount of shift of
$P(I_{SW})$ gets smaller and saturates, the rates $\tau ^{{\rm -}
{\rm 1}}(I_{SW})$ overlap with each other. This implies that $\tau
^{{\rm -} {\rm 1}}(I_{SW})$ is also independent of $T$ in this
regime. Using the value of $I_{C}$ obtained from the fitting of
$P(I_{SW})$, the tunneling rates are estimated from Eqs.
(\ref{eq1}) and (\ref{eq2}). The fitting of $\tau ^{{\rm -} {\rm
1}}(I_{SW})$ agrees with the experimental data.

\begin{figure}
\includegraphics[width=6.4cm]{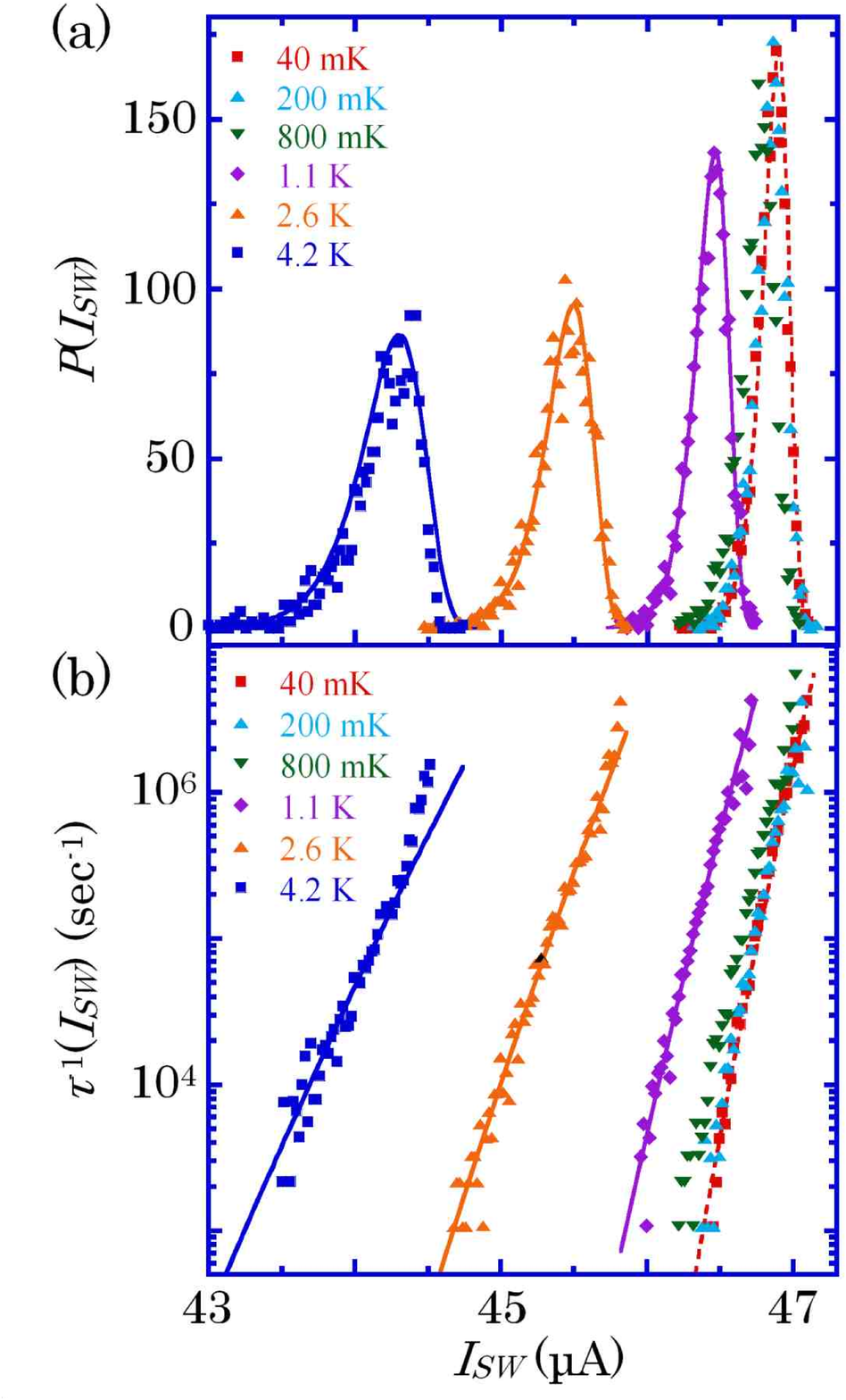}
\caption{\label{fig.2} (a) Switching current distribution $P$($I
_{SW}$) of IJJs. The solid and dashed lines are the
fitting \cite{Kramers:1940,Garg:1995} in the thermal
and MQT regions, respectively, and the zero-noise critical current
$I_{C}$ was extracted from these fittings. The measurement
condition and the junction parameters were the following:
$dI/dt$=42.4 mA/sec, $\Delta I=20$ nA, $I_{C}$=48.54$\pm $0.02 $\mu
$A, $C$=76.26 fF, $\omega _{p0}$/2$\pi $=221.3 GHz.
The junction capacitance $C$ was estimated by
$C=\varepsilon _{r}\varepsilon _{0}S/d$, where $d=15$ {\AA}
\cite{Kleiner:1992,Kleiner:1994} is the $c$-axis lattice constant
of Bi-2212, $S$ is the junction area, $\varepsilon _{r}=10$
\cite{Kleiner:1994} and $\varepsilon _{0}$ are the relative
permittivity and the vacuum permittivity, respectively. (b)
Escaping rate $\tau ^{{\rm -} {\rm 1}}(I_{SW})$ of IJJs. The
theoretical thermal rates calculated by Eq. (\ref{eq1}) are shown
as solid lines, and MQT rate without dissipation, which was
calculated by Eq. (\ref{eq2}) is exhibited as dashed line.}
\end{figure}

The standard deviation $\sigma (T)$ of $P(I_{SW})$ is plotted
against $T$ on a double logarithmic scale for two samples in Fig.
3. In the high $T$ region (1$\sim $4.2 K), $\sigma (T)$ is
proportional to $T^{{\rm 2}{\rm /} {\rm 3}}$ as expected, and
decreases with decreasing $T$ owing to the suppression of thermal
activation. But below 1 K it is saturated and independent of $T$
in accordance with Fig. 2. $T_{esc}$ \cite{Clarke:1988}
also exhibits a $T$-independent behavior. (see the inset of Fig. 3)
Actually, we have also observed this
kind of saturating behavior of $\sigma (T)$ in the IJJs fabricated
in Bi-2212 whisker single crystals. These phenomena agree with the
tendencies which have been reported for low-$T_{C}$ JJs
\cite{Voss:1981,Wallraff:2003}. To check that the observed saturation
is not due to external noise, we confirmed that connecting $LC$ low-pass
filters in series does not effect the value of $\sigma (T)$.
Moreover, an intermittent ramp bias was applied to the sample
in order to avoid heating. We are thus lead to conclude that
the experimental results presented here are direct consequences of
MQT of the macroscopic junction phase of IJJs fabricated in HTSC
Bi-2212. The effect of MQT appears below the crossover temperature
$T^{*}$, which we estimate from the data to be 0.75 K for
$I_{C}$=48.54 $\mu$A and 0.95 K for $I_{C}$=84.92 $\mu$A, which are
shown in Fig. 3 by the open and filled arrows, respectively. Meanwhile,
based on the uncertainties of $I_{C}$ and $C$, and the bias dependence of
$T^{*}$ as obtained from the formula $T^{*}\sim \hbar \omega _{p}/2\pi k_{B}$
\cite{Silvestrini:1997,Grabert:1984} is 0.75$\sim $0.80 K for
$I_{C}$=48.54 $\mu$A and 0.85$\sim $1.0 K for 84.92 $\mu$A.
These results agree well with the experimental values of $T^{*}$s.

\begin{figure}
\includegraphics[width=7.6cm]{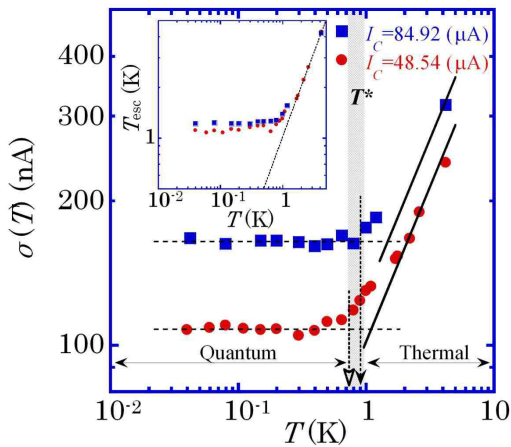}
\caption{\label{fig.3}Standard deviation $\sigma (T)$ of
$P(I_{SW})$. $\sigma (T)$ were calculated by $\sigma =(\langle
I_{SW}^{{\rm 2}} \rangle - \langle I_{SW} \rangle ^{{\rm
2}})^{{\rm 1}{\rm /}{\rm 2}}$ for two samples which had different
$I_{C}$s. The solid lines show the theoretical fitting of the
thermal region. $\sigma (T)$ starts to saturate below the
crossover temperature $T^{*}$
\cite{Silvestrini:1997,Grabert:1984}. The experimental $T^{*}$s of
the measured samples are around 1 K, falling within the shaded
crossover region calculated theoretically. The inset shows the escape
temperature $T_{{\rm esc}}$ vs. $T$. The dotted line corresponds to
$T$=$T_{{\rm esc}}$.}
\end{figure}

Having established the occurrence of MQT, let us briefly account
on its physical implications. We begin by mentioning two factors
essential to our findings, i.e. (\ref{eq1}) the large magnitude
$\omega _{p}$ of the overall energy scale involved, and
(\ref{eq2}) the relatively weak form of dissipation with which the
quasiparticle transport affects the system. The former aspect is
apparently reflected in the thermal-to-quantum crossover
temperature, which to the best of our knowledge is the highest
reported to date. Though the temperature itself is not the main
emphasis here, it does expose a potential advantage enjoyed by
HTSC materials, which deserves to be explored further. Meanwhile
the latter factor concerns the much more subtle issue of
``taming'' the nodal quasiparticles mentioned earlier, which bring
us back to the $I$-$V$ characteristics displayed in Fig. 1(c). In
general, the quasiparticle dynamics in a JJ is manifested in the
\textit{return curve} (the low-voltage subgap portion of the curve
following the jump to the first branch, obtained upon switching
off the bias current), a regime where Cooper pairs cannot
participate in the charge transport. In JJ stacks artificially
fabricated using conventional superconductors, this curve exhibits
an abrupt drop to nearly zero current at low temperatures
\cite{Kleiner:1994}, signalling a strong suppression of
quasiparticle excitations. Meanwhile, the
resistively-shunted-junction model widely employed in quantum
mechanical treatments of dissipation in JJs \cite{Caldeira:1981}
can lead to the dominance of ohmic behaviour and decoherence when
the resistance becomes a relevant perturbation. The power-law-like
tail seen in our data falls into a novel class of behaviour termed
the \textit{superohmic} dissipation \cite{Leggett:1987}, which is
intermediate between these two cases. That this subtle form of
dissipation should play a crucial role in our observations is in
accord with some of the recent theories on $d$-wave JJs
\cite{Fominov:2003,Amin:2004,Joglekar:2004,Kawabata:2004}.

Our study bores out a number of fundamental questions, which
reside on the intersection of two major areas of current research:
HTSC physics and future quantum technology. For example it
provides a fresh perspective on the issue of differentiating
between very distinct forms (e.g. transverse-momentum conserving
and nonconserving processes) of quasiparticle tunnelling between
$d$-wave superconductors, several of which may account for
\textit{superohmic} dissipation. Another related problem --with
direct implications to optimizing device geometry- is to resolve
how the differences between generic $d$-wave JJs and the
interlayer IJJs would influence MQT experiments. For instance, the
absence of zero-energy bound states \cite{Kawabata:2004}and the
presence of a bilayer tunnelling matrix element suppressing nodal
quasiparticle tunnelling \cite{Xiang:1996} in the latter can both
be considered advantages peculiar to the IJJs

In summary we have successfully observed phase MQT in the IJJs of
HTSCs, inviting further investigation on coherence-dissipation
competition in these systems. Though yet an initial step towards
applications, we anticipate that efforts along this line may lead
to a rich cross-fertilization between the fields of HTSC science
and quantum information.

We thank M. Kinjo for his technical assistance. This work was
supported in part by PRESTO, JST, and a Grant-in-Aid from the
Ministry of Education, Science, Sports, Culture and Technology,
Japan. S. K. was supported by NEDO-SYNAF.

\textbf{\textit{Note added- }}After completion of this work, we
learned that Bauch \textit{et al.} have also observed the MQT at
around 40 mK in an HTSC YBa$_{2}$Cu$_{3}$O$_{7-\delta }$ grain boundary
Josephson junction \cite{Bauch:2005}.

\end{document}